\documentclass{cimento}

\usepackage{graphicx}  

\title{Searching for New Physics with multilepton events at PADME}
\author{
G.~Martelli\from{ins:x}\thanks{Speaker: E-mail: gabriele.martelli@cern.ch}\ETC,
P.~Ciafaloni\from{ins:y}\thanks{E-mail: paolo.ciafaloni@le.infn.it}
\atque
M.~Raggi\from{ins:z}\thanks{E-mail: mauro.raggi@roma1.infn.it}
}
  
\instlist{\inst{ins:x} Dipartimento di Fisica e Geologia, Universit\`a di Perugia - Perugia, Italy
\inst{ins:y} INFN, Sezione di Lecce e Universit\`a del Salento - Lecce, Italy
\inst{ins:z} Dipartimento di Fisica, Universit\`a di Roma - Roma, Italy
}

\begin{document}

\maketitle

\begin{abstract}
The PADME experiment is searching for the Dark Photon $A'$ in the $e^{+}e^{-} \to \gamma A'$ process, assuming a $A'$ decay into invisible particles. In extended Dark Sector models, a Dark Higgs $h'$ can be produced alongside $A'$ in the process $e^{+}e^{-} \to h' A'$. If the $h'$ mass is greater than twice the $A'$ mass the final state will be composed by three $e^{+}e^{-}$ pairs. Such extremely rare process is explorable by the PADME experiment, which could get a first measure and impose limits on models of physics beyond the Standard Model.
\end{abstract}

\section{Introduction}
The extreme difficulty in Dark Matter detection could be explained speculating that Standard Model (SM) and Dark Matter particles exist in two separate sectors connected by a portal. The simplest model for this theory adds a new gauge group having $U_ {D}(1)$ symmetry which introduces a new boson $A'$, called Dark Photon. This new sector, and therefore the new boson introduced, is expected to couple with the SM electromagnetic field with a coupling constant $\epsilon$ of the order of $10^{-3}$ or smaller.\\
Many experiments in recent years have focused their attention on the production of this new vector boson, searching for both visible and invisible $A'$ decays. Among these lies the PADME experiment, which aims to measure processes $e^{+}e^{-} \to \gamma A'$ in the interaction of a positrons beam with the electrons of a diamond target, using the beam extracted from the DA$\Phi$NE linac at the National Laboratories of Frascati, LNF. The used technique is that of the missing mass $M^{2}_{miss} = (P_{e^{-}} + P_{beam} - P_{\gamma})^{2}$ in which only known kinematic variables are used. The presence of a vector boson $A'$ would manifest as a narrow peak in the spectrum of the variable $M^{2}_{miss}$ corresponding to its mass.\\
Although PADME was designed to the search for the $A'$, other Dark Matter candidates can be observed with this experiment. One of these is the the Dark Higgs ($h'$) which is introduced in models where the $A'$ mass is generated through the spontaneous symmetry breaking mechanism. This particle can be produced in PADME via the reaction $e^{+}e^{-} \to h' A'$. If $m_ {h '} \geq 2m_{A'}$ the Dark Higgs decays in a pair of $A'$, which in turn decays into a lepton pair, producing a six charged leptons final state, respectively three positrons and three electrons.\\
This article is focused on the search for Dark Higgs mediated six charged leptons, exploiting a theoretical study of the phenomenology of these low-energy events in the PADME experiment.\\

\section{The Dark Higgs}
In non-minimal models where the mass of the $A'$ is generated through spontaneous symmetry breaking, an associate production of a new boson, referred as Dark Higgs \cite{Batell:2009yf}, together with an $A'$ is possible. Naturalness requires that the two particles have masses of the same order $m_{h'} \sim m_{A'}$. This new sector is neutral under the Standard Model and vice versa, and all interactions with it proceed through kinetic mixing of $U_{D}(1)$ with the SM photon.\\
The Lagrangian containing the physical $h'$ field takes the form:

\begin{equation}
\mathcal{L}=-\frac{1}{4}A^2_{\mu \nu}+\frac{1}{2}m^2_{A'}+\frac{1}{2}(\partial_{\mu} h')^2++\frac{1}{2}m^2_{h'}h'^2+\mathcal{L}_{int}
\end{equation}

\begin{figure}[htb]
\begin{center}
\includegraphics[width=6.5 cm]{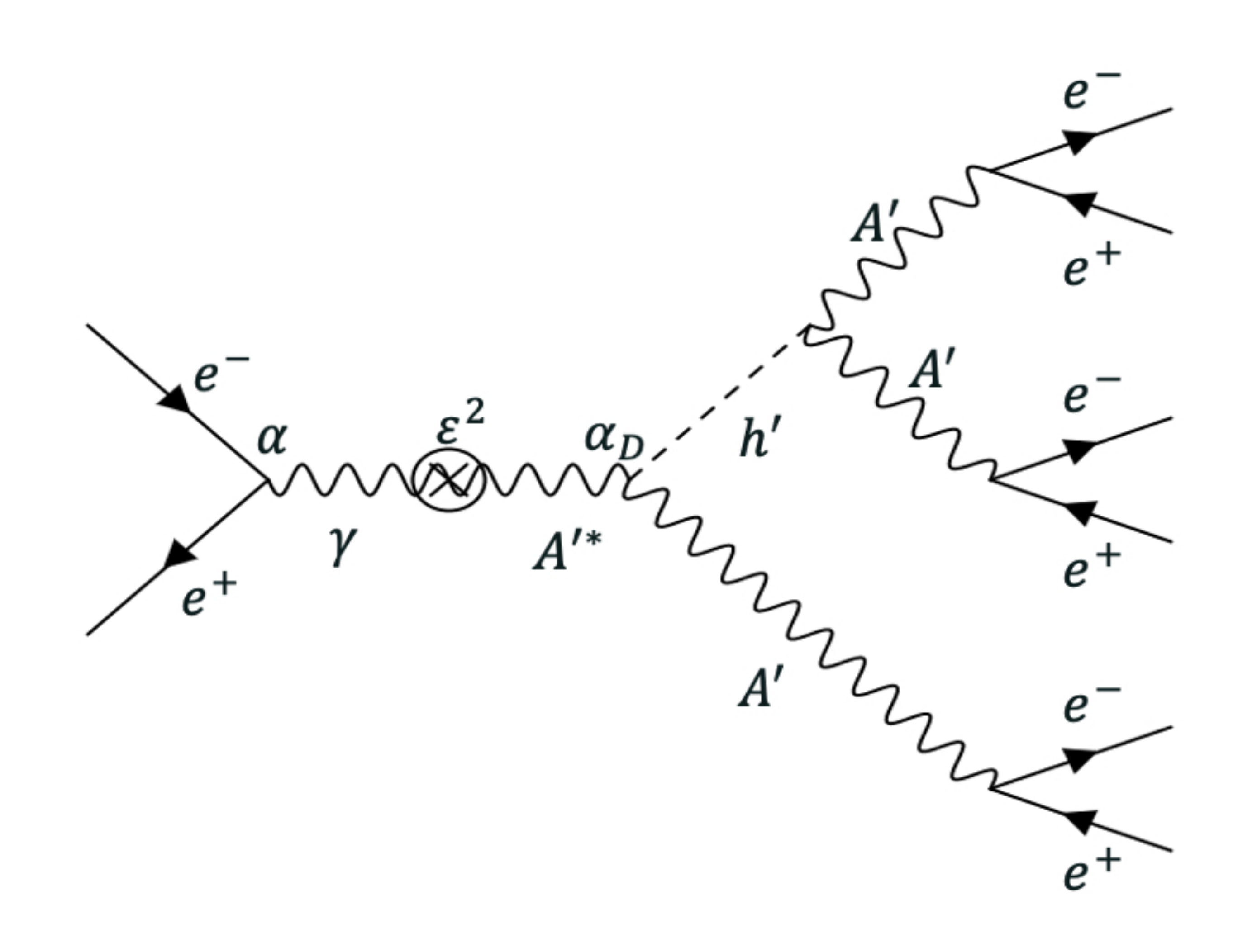}\\
\caption{{\em Feynman diagram for the Higgs'-strahlung process producing a six leptons final state.}}
\label{fig:hstrahalung}
\end{center}
\end{figure}

One of the few $h'$ production processes is the so-called Higgs'-strahlung, $e^{+}e^{-} \to h' A'$, which has an amplitude that is suppressed by just a single power of the kinetic mixing angle and can therefore readily occur for $\epsilon \sim O(10^{-2} - 10^{-3})$. This production mechanism is similar to SM Higgs-strahlung but in this case the $h'$ is produced in association with a $A'$ instead of a classic photon.\\
While the vector $A'$ will typically have a large branching ratio to lepton pairs, the decays of the $h'$ will depend on its mass relative to that of the vector. If the Dark Higgs is heavy it will decay in two vectors $A'$, eventually leading to a six lepton final state, as can be seen in Fig. \ref{fig:hstrahalung}.

On the contrary, if $h'$ is light, it will decay via loop processes to leptons and possibly hadrons. In this case the Dark Higgs is long-lived and will most likely become an undetected particle.\\
Experiments at $e^+e^-$ colliders with few \mbox{GeV} center of mass energy and very high luminosity, Babar \cite{babar} and Belle \cite{belle}, had investigated $e^+e^-\to 3(\ell^+\ell^-)$ setting stringent limits on the existence of the $h'$ for energy scales in the $\sim$ \mbox{GeV} range. No data are available for masses of the $h'<$1 \mbox{GeV} which can be probed by low energy fixed target experiments such as PADME \cite{Raggi:2014zpa}.

\section{Cross sections and acceptance evaluations}
In order to search for new physics with multi-lepton events in a low energy experiment such as PADME, a theoretical treatment is necessary.\\
The main background event to the Higgs'-strahlung is the SM $e^{+}e^{-} \to 3(e^{+}e^{-})$ process. The cross section for this process is very hard to calculate due to the high number of particle in the final states, producing several thousand of Feynman diagrams which contributes to the amplitude. An order of magnitude estimate of this cross section 
can be obtained by using the Leading Log Equivalent Photon Approximation (EPA) and the cross section $\sigma(\gamma \gamma \to e^+e^-e^+e^-)$ as proposed by \cite{Budnev:1974de}. As noted in \cite{CiafaloniMartelliRaggi}, this 
approximation leads a gross overestimate of the cross section values at low energies, due to accidental cancellation.

\begin{equation}
\sigma_{3(e^+e^-)}=\frac{\alpha^2} {6\pi^2}\sigma_{(\gamma \gamma \to 4e)} \left(\log ^4\left(\frac{s}{m^2}\right)+\textrm{A} \log ^3\left(\frac{s}{m^2}\right)+\textrm{B} \log ^2\left(\frac{s}{m^2}\right)+\textrm{C} \log \left(\frac{s}{m^2}\right)+\textrm{D}\right)
\label{eqn:6l_FullEPA}
\end{equation}

\begin{figure}[htb]
\begin{center}
\includegraphics[width= 13.5 cm]{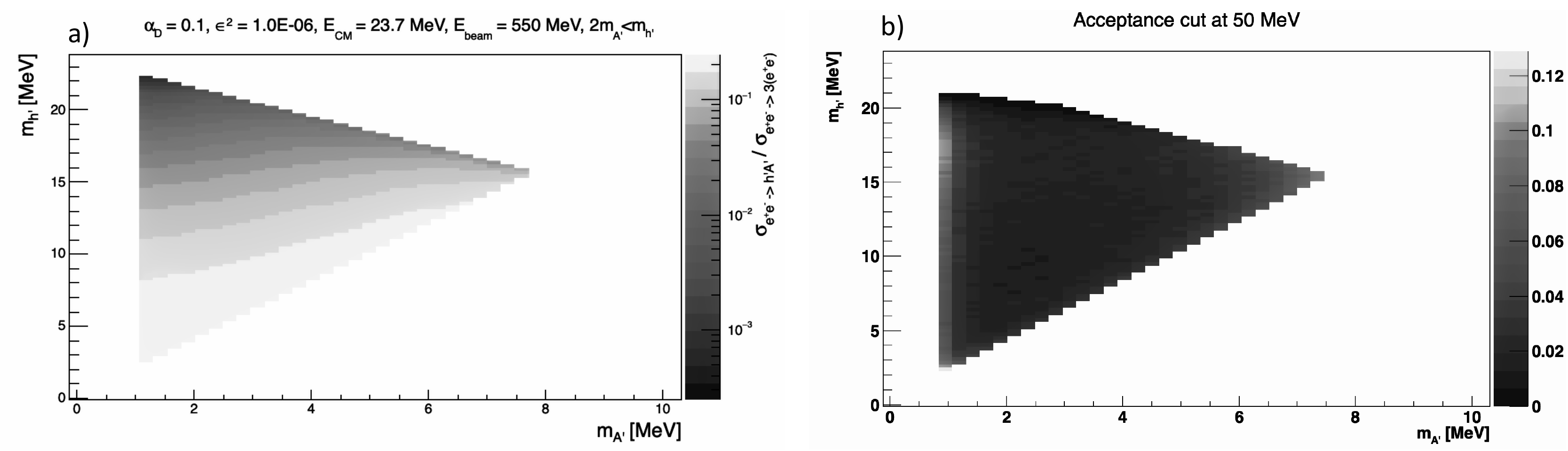}\\
\caption{{\em a) Ratio between the Dark Sector cross section and the Standard Model one. b) Acceptance values for the Dark Sector events at PADME.}}
\label{fig:Fig}
\end{center}
\end{figure}

Using Eq. \ref{eqn:6l_FullEPA} including all the logarithmic terms, at the PADME center of mass energy the full EPA estimate seem to point to a value $\sigma_{e^{+}e^{-} \to 3(e^{+}e^{-})} \sim 2000 $ \mbox{pb}, ten times smaller with respect to the leading log approximation.\\
The total cross section for the Higgs'-strahlung process reads \cite{Batell:2009yf}:

\begin{equation}\label{Eqn:DHCross}
\sigma_{e^{+}e^{-} \rightarrow A'h'} = \frac{\pi \alpha \alpha_D\epsilon^{2}}{3s} \left( 1 - \frac{m^{2}_{A'}}{s} \right)^{-2} \sqrt{\lambda \left( 1, \frac{m_{h'}^{2}}{s}, \frac{m^{2}_{A'}}{s} \right) }
\times \left[ \lambda \left( 1, \frac{m_{h'}^{2}}{s}, \frac{m^{2}_{A'}}{s} \right) + \frac{12m^{2}_{A'}}{s} \right] 
\end{equation}

where $\alpha_D$ is the coupling between $h'$ and $A'$, and $\lambda(a,b,c)\equiv a^2+b^2+c^2-2ab-2ac-2bc$.  
For accessible values of the kinetic mixing parameter $\epsilon$ $(10^{-4}\leq\epsilon\leq10^{-3})$, the cross section is quite large compared to the SM. In fact the Higgs'-strahlung cross section only pays an $\epsilon^2\alpha\alpha_D$ suppression compared to the $\alpha^6$ of the concurrent SM process. For fixed values of $m_{h'}$ and $m_{A'}$ the production cross section scales as 1/s.
In Fig. \ref{fig:Fig}a), assuming $m_{h'} + m_{A'} < \sqrt{s}$ and that $m_{h'} > 2m_{A'}$, the ratio between the possible cross section values of the Dark Sector process at PADME with the SM background is shown. The highest values of the Dark Sector cross section are obtained for low values of $m_{h'}$ and $m_{A'}$.\\
To evaluate the PADME spectrometer detection capabilities, a study of the acceptance has been made using a phase space simulation. Using ROOT's TGenPhaseSpace class and applying the kinematic limits explained above, it was possible to obtain the momenta distributions for the particles in the final state from which the acceptance was extracted, expressed in terms of $m_{h'}$ and $m_{A'}$. The PADME spectrometer is able to detect all of the six leptons, only if the minimum energy is $\ge$ 50 \mbox{MeV}, which is challenging having 6 tracks and 550 \mbox{MeV} total energy. The acceptance studies of the Dark Sector signal events are summarised in Fig. \ref{fig:Fig}b).

The acceptance is of the order of 1\% for most $h'A'$ pairs of masses.

\section{Conclusions}  
The theoretical studies presented in this article show that the Higgs'-strahlung process can also occur within the low-energy PADME experiment. However the acceptance study indicates a high efficiency of the PADME spectrometer for momenta higher than 50 \mbox{MeV}. The PADME experiment has therefore at present a low efficiency in reconstructing Dark Higgs events due to the high number of leptons. To increase the detection efficiency of low energy particles, the intensity of the magnetic field should be reduced, preventing low momentum charged particles from colliding with the vacuum chamber and increasing the detection efficiency of the first scintillating fingers of the veto systems.\\
Furthermore, the final state of these processes will be composed of a high number of charged particles. Therefore it is necessary to implement an efficient event reconstruction algorithm that can well identify the large number of charged particle pairs produced in the PADME spectrometer.

\end{document}